\documentclass[
twocolumn, 
prl,
superscriptaddress
]{revtex4-1} 

\usepackage{amssymb,amsmath}
\usepackage[utf8]{inputenc}
\usepackage{url}
\usepackage{hyperref}
\hypersetup{
    colorlinks=true,
    linkcolor=blue,
    filecolor=blue,      
    urlcolor=blue,
}
\usepackage{graphicx}
\usepackage{epstopdf}
\usepackage{subcaption}
\usepackage{color}
\usepackage{dcolumn}
\usepackage{bm}
\usepackage{comment}
\usepackage{ulem} %
\begin{document}

\title{Physics-informed Gaussian Process for Online Optimization of Particle Accelerators}

\author{Adi Hanuka}
\email{Corresponding author: adiha@slac.stanford.edu}
\author{X. Huang}
\author{J. Shtalenkova}
\author{D. Kennedy}
\author{A. Edelen}
\affiliation{SLAC National Accelerator Laboratory, Menlo Park, CA 94025, USA}
\author{V. R. Lalchand}
\affiliation{University of Cambridge, United Kingdom}
\author{D. Ratner}
\author{J. Duris}
\affiliation{SLAC National Accelerator Laboratory, Menlo Park, CA 94025, USA}

\date{\today}

\begin{abstract}
High-dimensional optimization is a critical challenge for operating large-scale scientific facilities. 
We apply a physics-informed Gaussian process (GP) optimizer to tune a complex system by conducting efficient global search. Typical GP models learn from past observations to make predictions, but this reduces their applicability to new systems where archive data is not available. 
Instead, here we use a fast approximate model from physics simulations to design the GP model. The GP is then employed to make inferences from sequential online observations in order to optimize the system.
Simulation and experimental studies were carried out to demonstrate the method for online control of a storage ring. We show that the physics-informed GP outperforms current routinely used online optimizers in terms of convergence speed, and robustness on this task.
The ability to inform the machine-learning model with physics may have wide applications in science. 
\end{abstract}

\maketitle

Online control and tuning of modern particle accelerators, such as free electron lasers and storage ring light sources, is a challenging task, since those systems often consist of hundreds of correlated parameters that could be adjusted in order to find a set of parameter values to achieve optimal target performance.
Automated tuning can help deliver the highest beam quality to scientific users during operation, and reduce tuning time for operation mode switching. This would be enabled by efficient online optimization algorithms, which are necessary in particle accelerators because, although physics models exist, there are often significant differences between the simulation and the real accelerator. The critical requirement for a suitable tuning algorithm is the ability to robustly find the optimum in a complex parameter space with high efficiency (minimum number of steps). 

Traditional model-independent optimization methods, that don't require the gradient of the system, such as Nelder-Mead simplex \cite{simplex}, may not work well for online applications when the target is noisy. 
Other local, model-independent methods, such as robust conjugate direction search (RCDS) \cite{rcds,rcds_demo} and extremum seeking (ES) \cite{ES} offer resilience to noise by taking a large number of samples, thereby often taking a long time to converge, or require some fair initial conditions \cite{ES:warmstart}. 
Machine learning (ML) model-based optimization methods may be beneficial to improve the quality of the solution, the speed of convergence and robustness to noise.

ML model-based methods for online optimization typically rely on learning from previously observed data. However, limited sparse sampling of high dimensional archived data may be insufficient, for example when learning correlations between various control variables. In addition, learning from archive data becomes impossible when preparing for new configurations where relevant experimental data does not exist. On the other hand, approximate physics models cannot be applied directly on the system to be optimized. Those need to be calibrated, and even then cannot exactly fit the observed data. In this Letter we circumvent the limitations of both approaches by approximating the covariance of the system directly from the physics model and then building a model from a few online observations. Physics models may capture the qualitative response of the objective with respect to controls better than archive data. Incorporating those into ML models may increase the speed of convergence and robustness of an online tuning process. 

Bayesian optimization is a model-based approach to optimizing expensive to evaluate, black-box systems with possibly noisy inputs and outputs \cite{mockus1975, freitas:bayes_opt_review, freitas:tutorial}.
Its effectiveness derives from probabilistic models of the system, such as Gaussian processes (GPs) \cite{rw:gpml_book}, which provide not only a prediction of the system's response, but the uncertainty in that prediction as well. GPs predict a distribution of possible functions compatible with observations by utilizing a covariance function, called the kernel, describing relationships between those observations. An attractive feature of GP modeling is the interpretability of the kernel’s functional form. The flexibility to capture the complex dependencies encountered in modern experiments lies in the design of this kernel.
Learning the kernel function rather than the target function itself is less prone to errors resulted from dependencies on drift or random hidden variables.

Recently, Bayesian optimization with Gaussian process surrogate models has been successfully demonstrated on linear accelerators ~\cite{mitch:spwogp, Kirschner_SwissFEL, Duris_lclsBO, Hanuka_SPEARBO}. Refs.~\cite{mitch:spwogp, Kirschner_SwissFEL} contain GPs with diagonal kernels (without correlations) learned from archive experimental data. 
%
In Ref. \cite{Duris_lclsBO}, we learned correlations from a physics model, but still required archived machine data to learn the length scales to build the full kernel. This was done in part because there was not a complete physics model available. 
The ability to easily learn the full kernel directly from a physics model would turn GPs into a practical tool applicable for tuning new machines and configurations without any archived data. 


In this Letter we experimentally demonstrate a {\it physics-informed} Bayesian optimization, where we use a physical model to directly derive the GP kernel including correlations. 
As an alternative to the traditional empirical kernel learning procedure using prior data \cite{rw:gpml_book}, we construct the kernel from the physical model's basis functions. The basis function kernel
eliminates the need for many data samples (either observed or simulated) and empirical kernel selection through marginal likelihood maximization (referred as ML-II).
As our primary result, we demonstrate experimentally the {physics-informed basis function} approach effectiveness by comparing performance with the traditional {data-informed ML-II} approach, and several other algorithms on the SPEAR3 storage ring \cite{spear3} facility for minimizing the vertical emittance with respect to 13 skew quadrupoles magnet. We finally discuss the importance of constructing a kernel and prior mean that is representative of the system to be modeled.

\textit{Methods.—}
Online tuning by Bayesian optimization involves two main components: (i) an online surrogate model $g(x)$ of how the objective $f(x)$ responds to a vector of input control values $x$ (e.g. beam loss rate with respect to 13 skew quadrupole magnet strengths). This model is iteratively updated with observed data during optimization. (ii) An acquisition function which chooses the next state based on the current state of the model built from the observed data.

The surrogate model we chose is a Gaussian Process (GP) \cite{rw:gpml_book} --- Bayesian non-parametric model which induces a prior over mean and covariance functions $g(x) \sim {\mathcal GP} (m(x),k(x_{\rm i},x_{\rm j}))$, where $x_{\rm i}, x_{\rm j}$ are all possible pairs in the input domain. The mean function $m(x)$ describes the expected value of the objective, and the kernel $k(x_{\rm i}, x_{\rm j})$ characterizes similarities between possible objective function values at different input points $x_{\rm i}$ and $x_{\rm j}$. While the optimum of the objective function may fluctuate day to day, the kernel captures the underlying behavior, allowing it to well represent the function given sampled data.

To account for the observations' noise, we model the noise as independent and identically distributed Gaussian random variables with a zero mean and a variance of $\sigma_{n}^2$. The corresponding Gaussian noise kernel is $k_{\rm noise}(x_{\rm i},x_{\rm j})= \sigma_{n}^2{\delta}_{\rm i,j}$, ${\delta}$ is the Kronecker delta function. 
The GP is constructed directly from sampled instances, thus allowing the model's complexity to grow with observations and adapt to previously unexplored regions of the input space.

One of the critical steps in achieving an operational GP optimizer for complex systems is constructing a kernel which encodes the underlying behaviour and relationships in the modeled data. 
%
For systems with complex high-dimensional data structures, expressive kernels facilitate efficient learning from online acquired data. Existing techniques to create expressive kernels from simpler ones include adding or multiplying kernels \cite{duvenaud:kernel_composition, sun:nkn} or applying a nonlinear transform to the input data \cite{wilson:dkl, manifoldgp, Deep-GP}.
In principle, general properties of kernels  are controlled by a number of hyperparameters. 

Usually, kernels and their hyperparameters are chosen by the type-II maximum likelihood method \cite{rw:gpml_book}. This ML-II method learns the hyperparameters of a chosen kernel which maximizes the marginal likelihood of historical data; see the Supplemental for more details on this approach~\cite{supplemental}. When using experimental archived data we refer to this approach as {\it data-informed ML-II} Gaussian process.
%
However, estimating a kernel's hyper-parameters from archive data becomes impossible  when preparing for new configurations. 

As an alternative, a physics simulation could be used instead of experimental data \cite{Yang:CoKriging,simGP:sensing}, making it possible to learn a kernel if there is only little or even no historical data at all. We refer to this approach to kernel construction as \textit{physics-informed}. 
However, as in the data-informed case, care must be taken in sampling the simulation input space to capture the objective's complexity as well as correlations between the input parameters.
Using simulation data can be expensive process and may require long computational time since high dimensional input space would require many evaluations of a possibly slow simulation. Then, using ML-II is costly, since the computational complexity scales as $n^3$ for $n$ data points. Therefore, there is a need to develop methods to find the best kernel and its hyperparameters without relying on many data samples (either observed or simulated), while allowing for the incorporation of prior physics knowledge. This would increase the kernel's interpretability, and may help gaining real insight into the system. 

In order to address this need, and to eliminate the requirement of empirical kernel selection using data (either observed or simulated), we calculate the kernel directly from a physical model.
%
There is growing interest in incorporating domain knowledge into kernel construction, including calculating the kernel directly from a physical model. For example, previous studies used governing partial differential equations to numerically calculate the covariance matrices \cite{PGPR:power_grid,NumericalKernel:turbulent,MultiGP:PDE}. 
In this work, we leverage the connection between infinitely wide Bayesian neural networks and Gaussian processes, to calculate the covariance function from an explicit basis function
\cite{MacKay98,neal:nn2gp,rw:gpml_book}:
\begin{equation}
    k({x_{\rm i}},{x_{\rm j}}) \propto \int_{-\infty}^{\infty} \phi({x_{\rm i}} - {c}) \phi({x_{\rm j}} - {c}) d{c} \label{eq:kernel_from_basis}
\end{equation}
where $c$ denotes the center of the basis function $\phi(x)$. We refer to GPs based on kernels designed in this way as {\it basis-function} GPs. Alternatively if the power spectral density (PSD) of the system is easier to model, the covariance function can be calculated from the amplitude of the Fourier transform of the PSD using the Wiener–Khinchin theorem \cite{wiener1930}.

For example, Radial Basis Function (RBF) of the form $\phi({x})=\exp(-x^{\rm T}\Sigma x/2)$, results in the RBF kernel function $
k(x_{\rm i},x_{\rm j})\propto \exp(-({x_{\rm i}}-{x_{\rm j}})^{T}(\Sigma/2)({x_{\rm i}}-{x_{\rm i}})/2)$ \cite{MacKay98,rw:gpml_book}, where $\Sigma$
is the precision matrix, and $(..)^T$ is the transpose operation. This type of basis functions are useful for modeling many smooth functions. The precision matrix is a symmetric matrix encoding properties of the function. For example, if there are no correlations between input parameters, $\Sigma={\rm diag}(l)^{-2}$ is a diagonal matrix wherein $l$ is a vector of characteristic length-scales. The latter specify how function values at two points separated in space along a single dimension (for example, a quadrupole magnet strength) relate to each other. 

\begin{figure} [htbp]
\centering
\begin{subfigure}[]{0.23\textwidth}
\centering
\includegraphics[width= 0.93\textwidth]{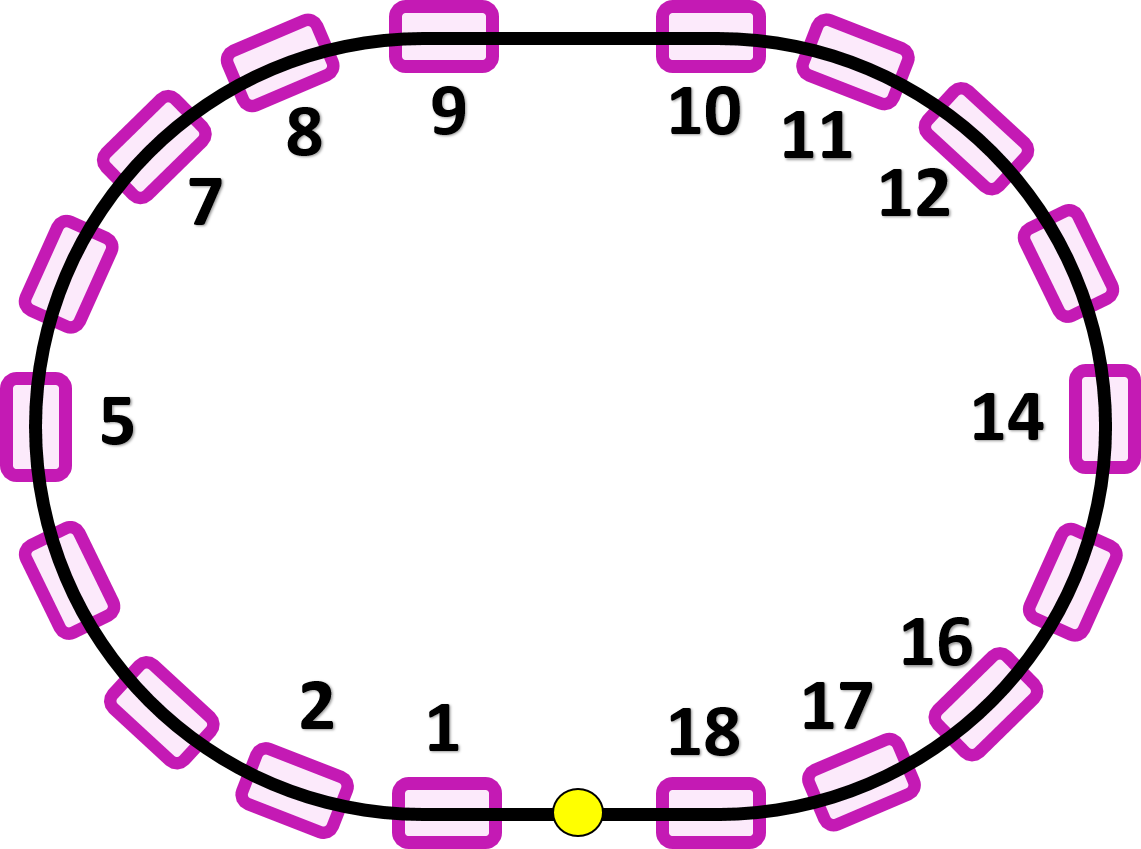}
 \caption{Layout of SPEAR3}
 \label{fig:layout}
\end{subfigure}
\begin{subfigure}[]{0.23\textwidth}
\centering
\includegraphics[width= 0.97\textwidth]{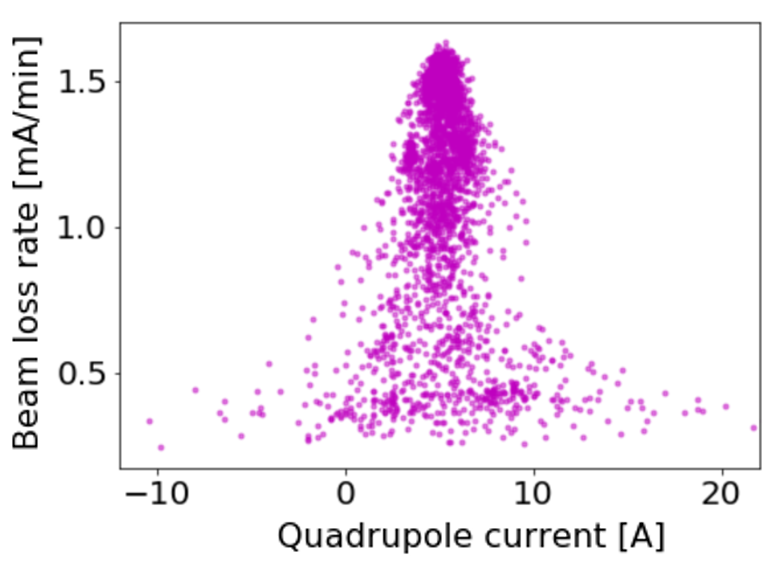}
 \caption{Beam loss rate}
 \label{fig:quadscan}
\end{subfigure}
 \caption{(\subref{fig:layout}) Layout of the SPEAR3 storage ring with the 13 free skew quadrupoles used for the online optimization of beam loss rate. The non-destructive current monitor is shown as a yellow dot. (\subref{fig:quadscan}) Beam loss rate projected on a single skew quadrupole current. The data is taken from archived operations scans.
 }
\end{figure}

In what follows, we use an approximation to the physics model as the basis function to design the kernel. This allows the GP to make predictions of the system using the covariance of the physics model as a estimate of that of the system. 
We refer to this approach as \textit{physics-informed basis-function} GP. 
Learning the kernel from simulated data instead of machine data is a form of kernel transfer learning \cite{KTL,KTL-CD}. 
Furthermore, constructing the kernel from basis functions without using the likelihood function is a form of Gaussian process with likelihood-free inference \cite{likefree, GPS-ABC}. 


%

In this Letter we consider the task of finding the peak of a system, which has a physics model of sufficient fidelity to capture the qualitative system's response \cite{good_regulator}. 
%
For example, the simulation could have an unknown scaling and translation with respect to the machine but its functional form is similar.  
In order to calculate the physics model's basis function, 
in this work, we consider systems that can be roughly approximated with a Gaussian around the optimum of the simulation. 
%
%
%
We then approximate the basis function by expanding the log of the simulation $\hat{f}(x)$ about a point $x_0$ close to the global optimum with an analytic expansion to second order, after subtracting off the asymptotic behavior $\hat{f}(\infty)>0$.
%

 %

%
We build the basis function by evaluating the gradient $(G)$ and the Hessian $(H)$ of the log of the simulation $G_{\rm i}=\partial_{x_{\rm i}}\log\big[\hat{f}(x)-\hat{f}(\infty)\big]|_{x=x_0}$, $H_{\rm i,j}=\partial_{x_ {\rm i}}\partial_{x_{\rm j}}\log\big[\hat{f}(x)-\hat{f}(\infty)\big]|_{x=x_0}$ via numerical differentiation. The resulting expansion is $\log\big[\hat{f}(x_0)-\hat{f}(\infty)\big] +(x-x_0)^{\rm T}G+\frac{1}{2} (x-x_0)^{\rm T} H (x-x_0)$. If the expansion point $x_0$ is an optimum as in the work presented here, then the gradient may be neglected, and the 
basis function has the functional form of a Gaussian:
\begin{equation}
\begin{aligned}
     \phi(x) =  [\hat{f}(x_0)-\hat{f}(\infty)\big]\exp\bigg[\frac{1}{2} (x-x_0)^{\rm T} H (x-x_0)\bigg].\label{eq:basis_fcn}
\end{aligned}
\end{equation}
%
%
Then Eq. \ref{eq:basis_fcn} is used to calculate the associated covariance function by applying Eq. \ref{eq:kernel_from_basis}. The resulting covariance function has the same functional form as the RBF kernel \cite{rw:gpml_book}, with a precision matrix half that of the Hessian above $\Sigma=-H/2$. 
%
The function value $\hat{f}(\infty)$ was taken into account as the GP prior mean. 

\begin{figure*} [!htp]
\centering
\begin{subfigure}[]{0.32\textwidth}
\centering
    \includegraphics[scale= 0.4]{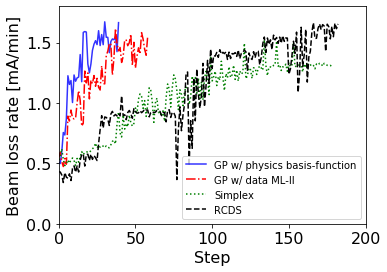}
    \caption{Online machine optimization - \\ Comparison of optimizers}
    \label{fig:live}
\end{subfigure}
\begin{subfigure}[]{0.32\textwidth}
\centering
    \includegraphics[scale= 0.4]{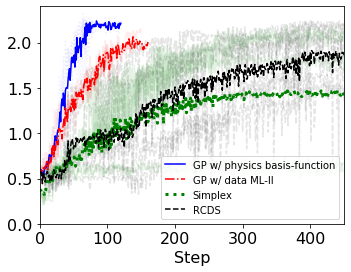}
    \caption{Simulated optimization - \\ Comparison of optimizers}
    \label{fig:sim}
\end{subfigure}
\begin{subfigure}[]{0.32\textwidth}
\centering
    \includegraphics[scale=0.4]{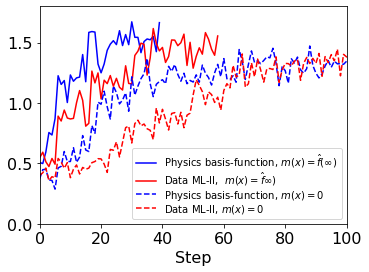}
    \caption{Online machine optimization - \\ Prior mean effect}
    \label{fig:live_bias}
\end{subfigure}
\centering
   \caption{(\subref{fig:live}) Comparison of optimization of beam loss rate over 13 skew quadrupole magnets for Gaussian process (GP) with physics-informed kernel including off-diagonal elements (blue), GP with diagonal-only data-informed kernel (red), Nelder-Mead simplex (green), and RCDS (black). Each step corresponds to approximately 1 to 2 seconds for GP and simplex, and to 6 seconds for RCDS.
   (\subref{fig:sim}) Simulations using the conditions of (\subref{fig:live}). Six individual scans for each method, with means shown by thick lines, are consistent with the {\it relative} performance of the online optimizations. 
   (\subref{fig:live_bias}) Comparison of GP optimizers with the objective's offset as prior mean (solid), and without (dashed).
    }
\label{fig:comparison}
\end{figure*}

\textit{Experiment.—}
In what follows we demonstrate experimentally the effectiveness of the physics-informed basis function approach on SPEAR3~\cite{spear3}, a third generation storage ring light source operating with low emittance, which results in high photon beam brightness.  
The goal of this optimization task is to minimize the average vertical emittance with skew quadrupoles. In an ideal electron storage ring, the vertical emittance is nearly zero. However, in reality there are various sources of errors that give rise to a finite vertical emittance, such as vertical dispersion in dipole magnets and linear betatron coupling between horizontal and vertical planes. Those error sources can be compensated by skew quadrupole magnets. In SPEAR3, there are 13 free skew quadrupoles for vertical emittance control (they do not change the horizontal emittance) - see Fig. \ref{fig:layout}.

Minimizing the vertical emittance is equivalent to minimize the vertical beam area.
Since the beam loss in the experiment is Touschek scattering dominated, minimizing the beam size corresponds to maximizing the beam loss rate (Amperes per minute) \citep{huang_2019_BBCOA}, which is non-disruptively monitored.  
We calculated the Hessian at the maximum beam loss rate point (see Fig. \ref{fig:quadscan}) in two ways. First, we used a fast-executing surrogate model trained on simulation data of the {\texttt{SPEAR3 storage ring Matlab Simulator}} \cite{rcds_demo} (details described in the Supplemental~\cite{supplemental}). This facilitates fast calculation of the Hessian. Second, we numerically calculated the Hessian directly from a noiseless SPEAR3 physics simulation. We found these to be in acceptable agreement.

%
%
While the precision matrix of the kernel containing both lengthscales and correlations was calculated from physics simulations, the kernel's amplitude was evaluated from the variance of a uniform distribution spanning the objective's range, which was similar to the value obtained by the ML-II. 
The kernel's noise was measured from a few live machine measurements $(\sigma_n=0.04)$. This noise is constant and is not correlated with the loss rate value.  



The skew quadrupole magnet currents were set to zero before scanning each time and this reduced the beam loss rate to $\hat{f}(\infty)\sim $0.5 mA/min.
The loss rate was evaluated by computing the change in the beam current loss rate over one second. Then we waited one second to let the quadrupoles current settle in the next point. For each set of experiments, the GP optimizer was initialized with a kernel, prior mean of $\hat{f}(\infty)$, and first observed point.

\textit{Results.—}
%
In what follows, we show that online optimization using the physics-informed basis function approach converges faster than the traditional data-informed ML-II approach. Since the basis-function approach makes it easier to calculate correlations, it is feasible to create a kernel including off-diagonal elements, whereas for the available data, the ML-II approach is limited to resolving a diagonal-only kernel.  
We also show that both methods surpass the current established optimization algorithms (Nelder-Mead simplex \cite{simplex} and robust conjugate direction search (RCDS) \cite{rcds}), which are routinely used to tune particle accelerator systems \cite{tomin:ocelot}. 


\autoref{fig:live} shows results from online optimization of the beam loss rate simultaneously on 13 skew quadrupole magnets. 
The GP optimizer with physics-informed basis function kernel reached an optimum of 1.67 mA/min in the smallest number of function evaluations (30 to 40 steps which are equivalent to 0.5 to 1 minutes).
The archive data-informed ML-II GP achieved 1.62 mA/min in 40 to 60 steps (0.66 to 1.2 minutes). The Nelder-Mead simplex optimizer achieved on 1.32 mA/min in approximately 160 steps (2.6 minutes). The RCDS optimizer achieved 1.66 mA/min, but took longer to converge; approximately 180 steps wherein each step is 6 seconds---total of 20 minutes. This increased measurement step time for RCDS allows for a reduced measurement noise of 0.02 mA/min which was found helpful for RCDS to converge. In contrast, the GP optimizers handle the noisier measurements better, resulting in shorter step times.

Although all optimizers, with the exception of Nelder-Mead simplex, found similar optimal loss rate within the measurement uncertainty (0.02 mA/min RMS for RCDS and 0.04 mA/min for the rest), the physics-informed basis-function found the optimum faster than even the data-informed ML-II GP, owing to the fact that it incorporates correlations between the quadrupoles to produce a better model. We also found that these results were consistent with subsequent tests where each optimizer started from the same \textit{random} starting point with the same initial beam loss rate as before.


A comparison of the above optimizers in simulation environment is shown in Fig. \ref{fig:sim}. Although the SPEAR3 simulator does not capture the full complexity of the live machine, it allows us to compare the \textit{relative} performance of the optimizers with a simulated objective function, which we find consistent with the online optimization. In simulation, on average, the physics-informed basis-function approach finds a better optimum in fewer iterations than the other methods. In addition, the spread of six individual scans for each method (with means in thick lines) reveals the robustness of the GPs, which follow similar trajectories for individual scans. 

Notably, the maximum available value of the simulated objective function is higher than the corresponding online optimization value.
This is understandable as the actual machine has more coupling error sources than modeled in the simulation. %
In reality, on the machine we cannot expect more than $\sim$1.7 mA/min with the experimental conditions used in this paper. 
Nevertheless, the simulation is adequate to capture the qualitative objective response with respect to skew quadrupoles in order to evaluate the physics-informed  covariance function. 

Our primary goal was to minimize the vertical emittance. In order to verify that, 
we performed the LOCO method \cite{Safranek1997} at three quads settings: the initial setting with 13 skew quads set to zero, the normal operation lattice, and the GP optimized solution. The emittance ratio between the vertical and horizontal planes 
was 0.71$\%$, 0.05$\%$, and 0.032$\%$ respectively.
Based on these combined simulation and experimental results, we expect the physics-informed basis function approach to be the most effective in practice.

%

Constructing a GP model that is representative of the system is a crucial step for increasing the effectiveness of the optimization. So far, we showed the effect of the kernel by comparing two GPs with a prior mean equal to the objective offset $(m(x)=\hat{f}(\infty))$. In what follows, we demonstrate the effect of the prior mean by comparing a GP model with a prior mean to one without $(m(x)=0)$.
Figure \ref{fig:live_bias} shows a comparison of the optimization results of the GP approaches with $m(x)=0$ and $\hat{f}(\infty)$. For both cases, the GP optimizers with zero prior mean (dashed lines) converged slower to a lower optimum, further validating the importance of the prior mean choice. In addition, the physics-informed basis function GP optimizer converged faster than the data-informed ML-II GP optimizer in both cases.

\textit{Conclusion.—} 
We presented and experimentally demonstrated a method incorporating physics models directly into a Gaussian process (GP) optimizer. Our method presents a simple way to construct the GP kernel, including correlations between devices. The physics-informed GP, which is more representative of the system, performed faster in an online optimization task compared to routinely used optimizers. 


In general, ML model-based methods for online optimization typically rely on many data samples. On the other hand, physics abounds with well verified mathematical models which we can exploit to learn approximate system dynamics in order to optimize new systems, without prior data. We computed the kernel using an approximated basis function from a physics model rather than from data samples.  
This method is faster to construct the full kernel, and could be easily adapted to other systems. It would also be applicable for automatic tuning and control of new machines and other complex configurations where historical data is unavailable or insufficient to resolve the kernel's hyper-parameters, including correlations. 
%
The basis function method is particularly well suited to analytical or differentiable models \cite{autodiffsim}, as well as surrogate models \cite{SM:AWA}.
%
The incorporation of prior physics knowledge would increase the attractiveness of Bayesian optimization with GPs for practitioners across various scientific domains, and may have wide applications in science. 
%

\begin{acknowledgments}
The authors are grateful to the SPEAR3 operators and engineers for their help with live tests on the storage ring.
This work was supported by the Department of Energy, Laboratory Directed Research and Development program at SLAC National Accelerator Laboratory, under contract DE-AC02-76SF00515, and by Office of Advanced Scientific Computing Research under FWP 2018-SLAC-100469ASCR.
\end{acknowledgments}


\bibliographystyle{apsrev4-1}
%

\end{document}